# Columnar grain boundaries are the weakest link in hard coatings: Insights from micro-cantilever testing with bridge notches

Yinxia Zhang[1], Matthias Bartosik[2], Steffen Brinckmann[3], Ujjval Bansal[1], Subin Lee[1,*], Christoph Kirchlechner[1]

[1]Institute for Applied Materials, Karlsruhe Institute of Technology, D-76131 Karlsruhe, Germany

[2]Department of Materials Science, Montanuniversität Leoben, A-8700, Leoben, Austria

[3]Microstructure and Properties of Materials (IMD-1), Forschungszentrum Jülich GmbH, D-52428, Jülich, Germany

*Corresponding author: Subin Lee, subin.lee@kit.edu*

**Abstract**

The effect of columnar grain boundaries on the fracture toughness was investigated using micro-cantilever fracture testing with a bridge notch in a unique hard coating system consisting of two distinct microstructures: one with columnar grains and the other with an epitaxial layer. The bridge-failure sequence qualitatively demonstrated the lower fracture toughness in the columnar-grained structure. Quantitatively, the load drops measured at bridge-failure also revealed a significant decrease in fracture toughness due to grain boundaries. Fracture toughness decreased by approximately 30%, from $4.1 \pm 0.4$ MPa m$^{1/2}$ for epitaxial microstructure to $3.0 \pm 0.3$ MPa m$^{1/2}$ for columnar-grained structure. These findings highlight the importance of grain boundary toughening in optimizing hard coatings and their mechanical properties.

*Keywords: Hard coatings, Grain boundary, Epitaxial structures, Fracture toughness*

Hard coatings are widely used for various industrial applications including cutting tools, protective layers for aerospace components, and wear-resistant surfaces for mechanical parts [1-3]. Typically composed of oxides, nitrides, or carbide, these coatings offer superior hardness, wear resistance, and thermal stability compared to metallic substrates [4-6]. However, a critical limitation of these hard coatings is their fracture toughness, which measures a material's ability to resist crack propagation. Hard coatings, particularly those prepared using physical vapor deposition (PVD) processes, often exhibit a crystallographic texture characterized by sub-micron sized columnar grains [7, 8] and a high density of grain boundaries (GBs). These GBs can potentially act as sites for crack initiation and propagation, thereby reducing the overall fracture toughness of the coatings [9-15]. In addition, such textured microstructures may introduce anisotropy in fracture behavior [16].

Previous studies have explored the influence of GBs on the mechanical properties of hard coatings [17-19]. For instance, the fracture behavior of α-$Al_2O_3$ coatings was investigated through micro-cantilever bending tests, which reported slightly enhanced fracture toughness in single crystal coatings compared to polycrystalline ones [19]. Conversely, improvements in mechanical properties through GB engineering have also been reported in zirconia coatings [18]; the polycrystalline samples showed better crack resistance under nanoindentation compared to a single crystal. Despite these findings, quantitative investigations on the effect of columnar GBs on the fracture toughness remain limited. One of the challenges is that various factors, besides GBs, can influence the (apparent) fracture toughness of hard coatings, such as residual stress, off-stoichiometry, and chemical inhomogeneity. These intrinsic properties are strongly affected by the deposition process, where even small variations in the deposition parameters can significantly change the properties of the coatings [20, 21]. Consequently, it is extremely challenging to synthesize two model systems with identical or similar chemical/defects structures but differences only in their grain size. This complexity hinders clear and fair comparison between coatings with and without GBs, and makes it difficult to gain quantitative insights into the effect of columnar GBs on the fracture toughness.

To address this gap, this study investigates the influence of GBs on the fracture toughness of hard coatings with a model system of CrN/AlN multilayered hard coatings. Unique to the presented approach is the integration of two distinct microstructures within a single coating: one exhibiting a columnar-grained microstructure typical for PVD, and the other with epitaxial film growth. This unique coating together with *in situ* scanning electron microscopy (SEM) micro-cantilever fracture testing using a bridge notch geometry designed to arrest cracks, enables discrete measurements of the fracture toughness of the columnar-grained and the GB-free epitaxial thin film, thus isolating the effects of columnar grain boundaries on the fracture toughness.

CrN/AlN multilayered coatings, consisting of alternating layers of nominally 4 nm CrN and 2 nm of AlN, were deposited on MgO (100) and Si (100) substrates [22]. Prior to deposition, the substrates were ultrasonically cleaned in acetone and ethanol, respectively. They were then mounted in an AJA ATC-1800 ultra-high vacuum deposition system, where they underwent thermal cleaning at 550°C for 30 minutes in a vacuum, followed by 5–10 minutes of Ar ion etching at 500°C. The Cr (diameter three-inch, purity 99.95%) and Al (diameter three-inch, purity 99.99%) targets were sputter-cleaned behind closed shutters for 5 minutes before film growth. The coatings were subsequently grown in pulsed DC mode (100 kHz pulse frequency, 1 µs pause). To ensure a dense morphology, 300 W was applied to the Cr target and 500 W to the Al target in a mixed $N_2$/Ar gas atmosphere (12 sccm / 8 sccm flow rate ratio) at a total pressure of 0.2 Pa, combined with a -70 V DC bias applied to the substrates. The substrates were continuously rotated at approximately 0.5 Hz during deposition. The multilayered structure was achieved through computer-controlled opening and closing of mechanical shutters at specific intervals. The total thickness of the CrN/AlN multilayered coatings was 1.9 μm.

Firstly, the microstructure of the coating was characterized using transmission electron microscopy (TEM, Titan Themis 300, Thermo Fisher Scientific). The multilayered coating on a Si (100) substrate exhibited a columnar-grained structure, whereas the coating on a MgO substrate had two distinct microstructures: an epitaxial structure at the bottom—roughly 500 nm thick—followed by a columnar-grained microstructure above, with a column diameter of 70–100 nm. The columnar-grained microstructure in the upper regions of the coating on MgO (shown in Figs. 1a and b) appears similar to that of the coating on the Si substrate (Fig. S1 in the Supplementary materials). Hereafter, the coating on Si will be referred to as the columnar grain (cg) coating, and the one with both epitaxial (epi) and columnar-grained microstructures as the cg/epi coating. Selected area electron diffraction patterns from each part of the cg/epi coating showed the (100) cube-on-cube orientation relationship between the MgO substrate and the epitaxial part, which is largely maintained in the upper part as well (Figs. 1c–e). High-angle annular dark-field (HAADF) scanning transmission electron microscopy (STEM) imaging and energy dispersive spectroscopy (EDS) mapping reveal the layered microstructure of the coating (Figs. 1f–h). Faulted layers as indicated by an arrow in Fig. 1f, show less dark contrast compared to other layers suggesting lower atomic density. Further, detailed comparisons of the cg coating and cg/epi coating, including microstructure analysis via the HAADF-STEM images, electron diffraction, and composition analysis via STEM-EDS, are provided in Fig. S2 in the Supplementary materials.

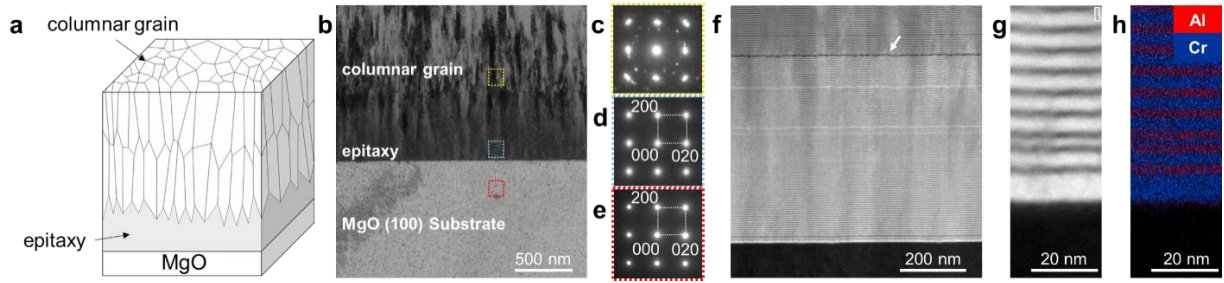

**Fig. 1** (a) Schematic drawing of the cg/epi-coating with an epitaxial layer between the columnar-grained microstructure and the MgO (100) substrate. (b) TEM image of the cross-section of the cg/epi-coating. Selected area diffraction patterns from (c) the columnar-grained microstructure, (d) the epitaxial structure, (e) and the MgO (100) substrate. (f) HAADF-STEM image of the epitaxial part. (g) High resolution of HAADF-STEM image on the epitaxial part and (h) its corresponding EDS maps.

Subsequently, the fracture toughness of the coating was tested using micro-cantilevers with square cross-sections fabricated using focused ion beam (FIB, Crossbeam 550L, Zeiss) milling with 30kV $Ga^+$ ions and ion currents of 15 nA, 3 nA, 700 pA, and 50 pA for step-wise milling. The dimensions of the micro-cantilevers were kept consistent with an *L:W:B* ratio of 5:1:1 shown in Fig. 2a, where $L$ is the distance to the loading point from a notch, $W$ is the thickness, and $B$ is the width of the micro-cantilever. A bridge notch with a depth, $a$, ranging from 20–30% of the thickness $W$ was introduced using a 20 pA and 30 kV ion beam. The width of the notch, $b$, was kept with a $b/B$ ratio of 0.92 to observe bridge-failure with subsequent crack arrest [22]. Micro-fracture experiments were performed *in situ* using a nanoindenter (Hysitron PI-89 NG, Bruker) equipped with a 10 μm diamond wedge tip (Synton-MDP AG) in an SEM (Merlin Gemini II, Zeiss). All tests were conducted in displacement-controlled mode at 5 nm/s using a transducer with a maximum load of 10 mN and a noise floor of 0.4 μN.

The influence of the epitaxial structure on the fracture toughness of the coating was investigated by comparing fracture toughness of cg/epi and cg-coating. In both samples, the load-displacement curves, 8 curves from cg/epi-coatings (Fig. 2b) and 11 curves from cg-coatings [22], exhibited linear elastic behavior until catastrophic brittle fracture.

As a first attempt, we analyzed the fracture toughness, $K_{IC}$, using the final failure of the specimens at point C (see Fig. 2b). Note that by this point, the two bridges had already failed, and the final fracture occurred with the arrested natural crack. This protocol follows Matoy's pioneering work introduced in [23],

$$K_{IC} = \frac{F_C L}{BW^{\frac{3}{2}}} f_{Matoy}\left(\frac{a}{W}\right) \tag{1}$$

where, $F_C$ is the load at final fracture, and $f_{Matoy}\left(\frac{a}{W}\right)$ is a geometry shape factor defined by:

$$f_{Matoy}\left(\frac{a}{W}\right) = 1.46 + 24.36(a/W) - 47.21(a/W)^2 + 75.18(a/W)^3 \tag{2}$$

The fracture toughness obtained at point C was statistically analyzed, and its cumulative distribution (Fig. 2c) clearly demonstrates that $K_{IC}$ of cg/epi-coating is higher than that of cg-coating: the mean $K_{IC}$ value is $3.1 \pm 0.1$ MPa m$^{1/2}$ for cg/epi-coating while $2.7 \pm 0.1$ MPa m$^{1/2}$ [22] for cg-coating with the error representing the standard deviation. The shaded band in Fig. 2c indicates 95% confidence intervals. Although the epitaxial part was located at the bottom of the coating and comprised only a quarter of the total thickness, it notably enhanced the apparent fracture toughness by nearly 15%. A comparison of the fracture surface reveals a flatter morphology in the epitaxial region adjacent to the MgO substrate (refer to Figs. 2d and e), and a surface roughness suggesting intergranular fracture in the cg as well as the upper portion of the cg/epi coating. These findings strongly indicate that GBs have a negative impact on the fracture behavior of hard coatings.

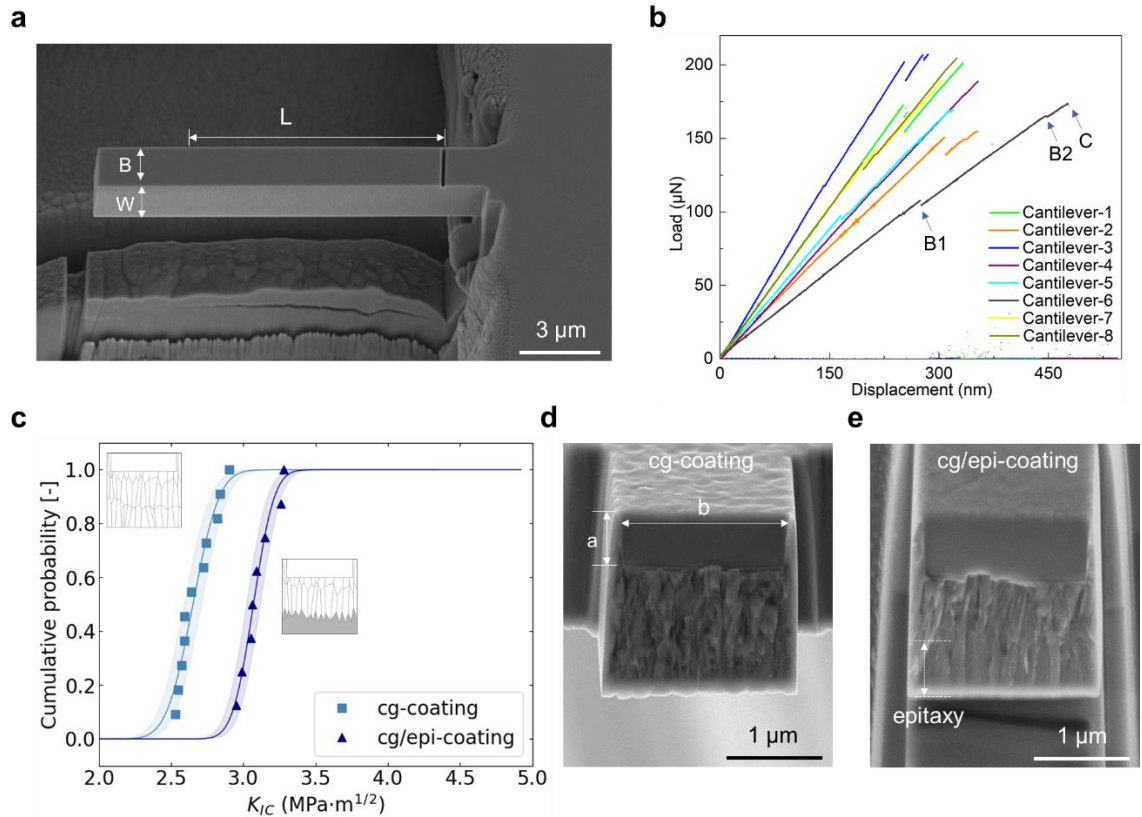

**Fig. 2** (a) Free-standing micro-cantilever of the cg/epi-coating. (b) Load-displacement curves from 8 micro-cantilever bending tests of the cg/epi-coating. The failure of the material bridges, B1 and B2, and the final fracture, C, are indicated on one representative curve. (c) Cumulative distribution of the fracture toughness for the cg/epi-coating and cg-coatings. The cg/epi-coating with epitaxial structures exhibited higher fracture toughness. (d) SEM image of the fracture surface of the cg-coating and (e) the cg/epi-coating. Intergranular fracture is visible in the cg-coating and the top part of the cg/epi-coating.

For distinct determination of the apparent fracture toughness of the columnar-grained and the epitaxial microstructure, we employed bridge notches in 90-degree rotated micro-cantilevers (compare the growth directions in Fig. 3a). This approach allows for positioning each material bridge in either of the two microstructures. The bridge-failure sequence, i.e., the temporal occurrence of failure at either the epitaxial or the columnar microstructure as presented in Fig. 3b, already indicates a higher toughness of the epitaxial layer: the columnar-grained microstructure (left bridge) failed first in the cg/epi-coatings, while a uniform distribution was found for the cg coatings, indicating a clean load with negligible misalignment of the wedge indenter. The change in failure sequence in the cg/epi sample clearly demonstrates differences in the fracture behavior of the epitaxial and columnar microstructures, and qualitatively indicates that columnar grain boundaries are indeed the weakest link in hard coatings, exhibiting lower fracture toughness compared to the epitaxial microstructure.

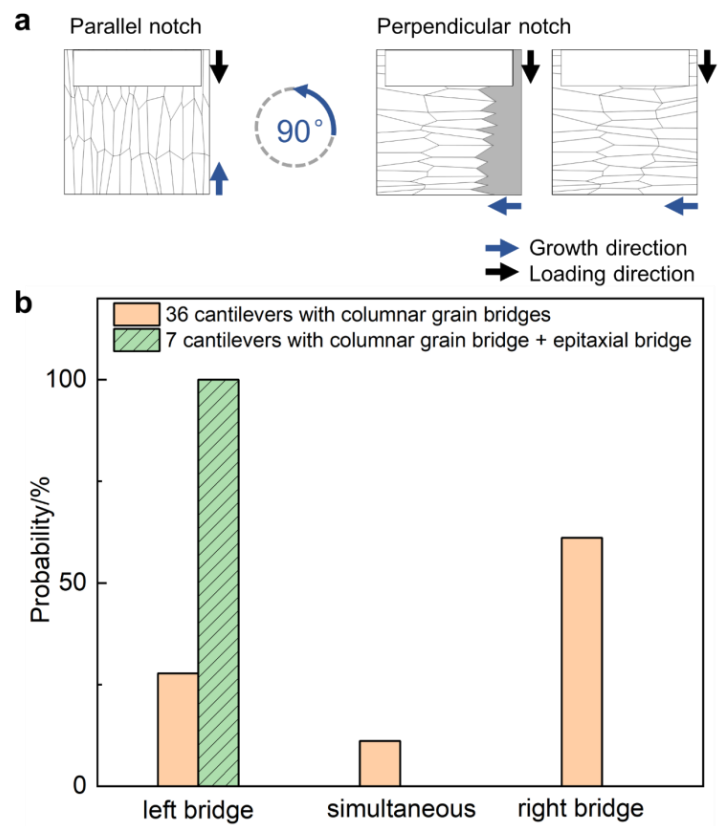

**Fig. 3** (a) The schematics showing cross-sections of micro-cantilevers with bridge notches. Material bridges were positioned within each microstructure, i.e., the columnar grain and epitaxial microstructure, by rotating the coating 90 degrees. (b) Statistics of bridge-failure sequence. The orange bars represent 36 micro-cantilevers only with cg-bridges, and the green bar shows 7 micro-cantilevers from the cg/epi-coating with a cg-bridge on the left and an epi-bridge on the right side. The failure sequence is random when the microstructure of both bridges is similar, while in the cg/epi-coating, the bridge with epitaxial microstructure was always broken after the columnar one.

Besides the failure sequence, we also used the load drops (B1 and B2 in Fig. 2b) caused by bridge-failure to compute the local apparent fracture toughness $K_{IC}^*$ from bridge-failure using the geometry correction factor $f_{corr}$ [22-24]:

$$K_{IC}^* = \frac{F_B L}{B W^{\frac{3}{2}}} f_{Matoy}\left(\frac{a}{W}\right) / f_{corr} \tag{3}$$

where, $F_B$ represents the load at bridge-failure.

In brittle ceramics, extrinsic toughening is often considered the primary mechanism for enhancing toughness, as discussed in Lawn's book [25]. For example, the fracture toughness of SiC is typically 2–3 MPa m$^{1/2}$ during transgranular fracture, but can increase to as much as 10 MPa m$^{1/2}$ during intergranular fracture [26]. Given this, one might expect significantly higher fracture toughness in the columnar microstructure than in the epitaxial layer. However, the epitaxial bridge showed a fracture toughness (denoted $K_{IC}^*$-$B_R$) of 4.1 ± 0.4 MPa m$^{1/2}$, which is higher compared to that of the columnar-grained bridge ($K_{IC}^*$-$B_L$), which is 3.0 ± 0.3 MPa m$^{1/2}$ (Fig. 4a). The latter value is consistent with experiments on a cg-coating in similar geometry i.e. with crack propagation perpendicular to the growth direction (see the orange markers in Fig. 4a, 3.0 ± 0.2 MPa m$^{1/2}$) as well as with the mean fracture toughness of the cg/epi-coating analyzed from the final failure (compare the purple markers in Fig. 2e, 3.1 ± 0.1 MPa m$^{1/2}$).

In addition to these quantitative measurements, the fracture surfaces also revealed clear differences between the epitaxial and the columnar microstructures. While the fracture surface is rough across the cross-section in the case of cg-coating (Fig. 4b), indicating intergranular fracture, the right side of the fracture surface of the cg/epi-coating was smooth, characteristic of intragranular fracture in the epitaxial microstructure (Fig. 4c). Further images of the fracture surfaces are shown in Fig. S3 in the Supplementary materials.

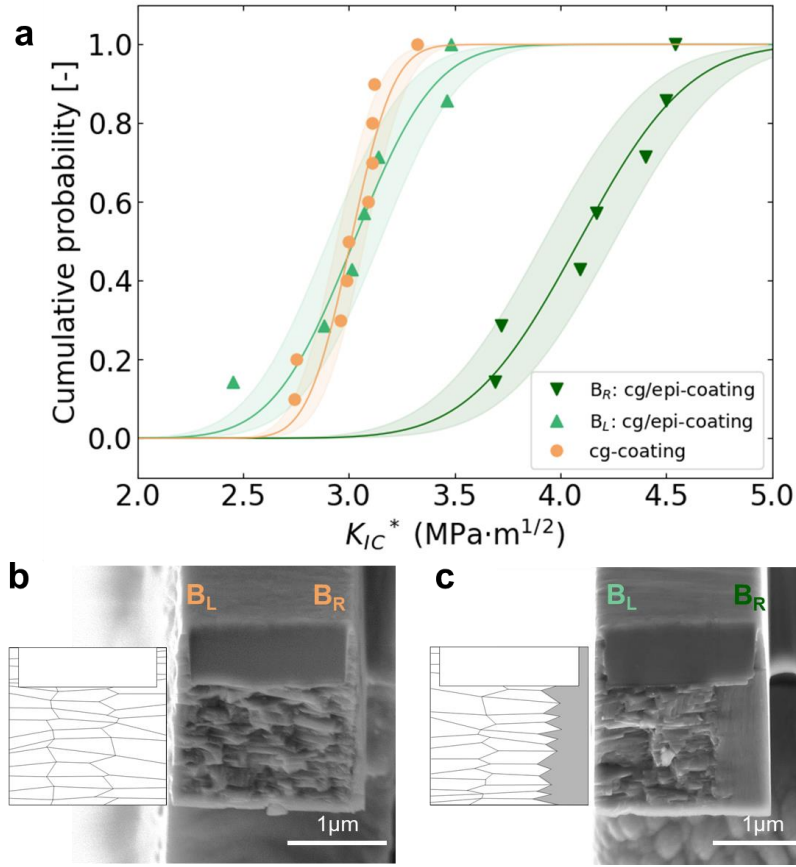

**Fig. 4** Comparison of the apparent fracture toughness from bridge-failures of two microstructures, and their fracture surface. (a) The cumulative distribution functions of the fracture toughness $K_{IC}^*$ derived from bridge-failure of the cg-coating, and the cg/epi-coating with crack growth perpendicular to the film growth direction. (b) SEM images of the fracture surface of the cg-coating and (c) the cg/epi coatings.

In conclusion, columnar grain boundaries clearly act as the weakest link in hard coatings, limiting their resistance to crack growth. This grain boundary weakness was demonstrated both qualitatively through the failure sequence of micro-cantilevers with a bridge geometry, and quantitatively through fracture toughness measurements at bridge-failure, using samples with two distinct microstructures: columnar grains and an epitaxial seed layer. The apparent fracture toughness measured from the bridges composed solely of the epitaxial microstructure was found to be around 30% higher compared to that of the columnar grains (compare $4.1 \pm 0.4$ MPa m$^{1/2}$ and $3.0 \pm 0.3$ MPa m$^{1/2}$). This clearly demonstrates that the columnar grain boundaries of the PVD hard coatings reduce fracture toughness. These findings suggest that future toughness optimization strategies and microstructure development in hard coatings could benefit from focusing on grain boundary toughening concepts.

**CRediT authorship contribution statement**

**Yinxia Zhang:** Writing- Original draft preparation, Methodology, Investigation, Formal analysis, Data curation, Visualization. **Matthias Bartosik:** Resources, Methodology, Writing- review & editing. **Steffen Brinckmann:** Writing- review & editing. **Ujjval Bansal:** Writing- review & editing, Investigation. **Subin Lee:** Writing- review & editing, Supervision, Software, Project administration. **Christoph Kirchlechner:** Writing- review & editing, Conceptualization, Supervision, Project administration, Funding acquisition.

**Declaration of Competing Interest**

The authors declare that they have no known competing financial interests or personal relationships that could have appeared to influence the work reported in this paper.

**Acknowledgement**

This research was funded within the framework of the DACH program by the national funding agencies: Austrian Science Fund (FWF) [I4720] and German Research Foundation (DFG) [436506789]. Support from the Helmholtz Program Materials Systems Engineering and from the Robert-Bosch-Foundation is gratefully acknowledged. The authors thank M. T. Becker for the film deposition, the Karlsruhe Nano Micro Facility (KNMFi) for support and access to TEM facilities and R. Pippan from the Austrian Academy of Sciences for helpful discussions.

# Columnar grain boundaries are the weakest link in hard coatings: Insights from micro cantilever testing with bridge notches

Yinxia Zhang[1], Matthias Bartosik[2], Steffen Brinckmann[3], Ujjval Bansal[1], Subin Lee[1*], Christoph Kirchlechner[1]

[1]Institute for Applied Materials, Karlsruhe Institute of Technology, D-76131 Karlsruhe, Germany

[2]Department of Materials Science, Montanuniversität Leoben, A-8700, Leoben, Austria

[3]Microstructure and Properties of Materials (IMD-1), Forschungszentrum Jülich GmbH, D- 52428, Jülich, Germany

*Corresponding author: Subin Lee, subin.lee@kit.edu*

# Supplementary Information

**Microstructure of the cg and the cg/epi coating**

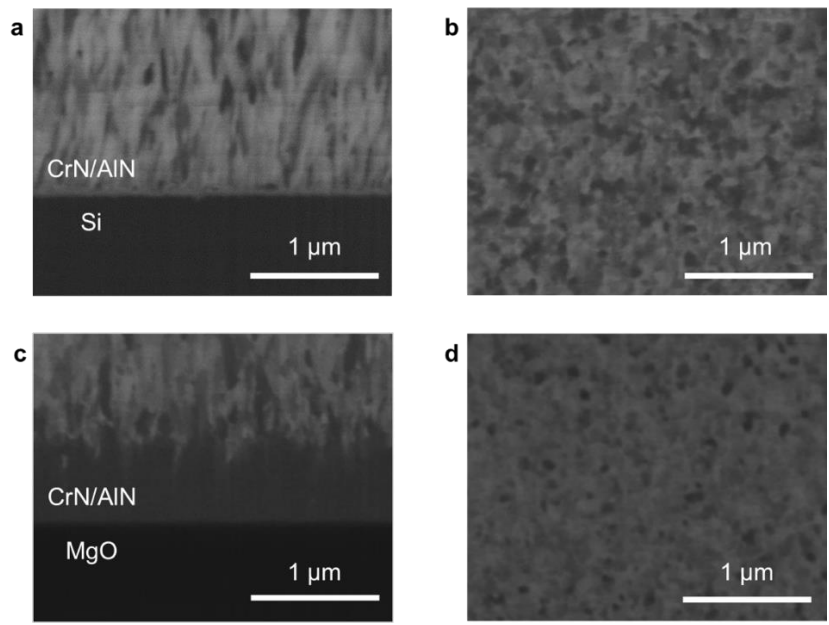

**Fig. S1** FIB channeling contrast images showing the microstructure of the coatings on Si and MgO substrate. On the left column present cross-section images, while the right column shows are top-view images. (a) and (b) Microstructure of CrN/AlN on Si (100) (cg-coating). (c) and (d) Microstructure of CrN/AlN on MgO (100) with an epitaxial layer (cg/epi-coating). The columnar grain structure in the upper regions of the coating on MgO appears similar to that of the coating on the Si substrate.

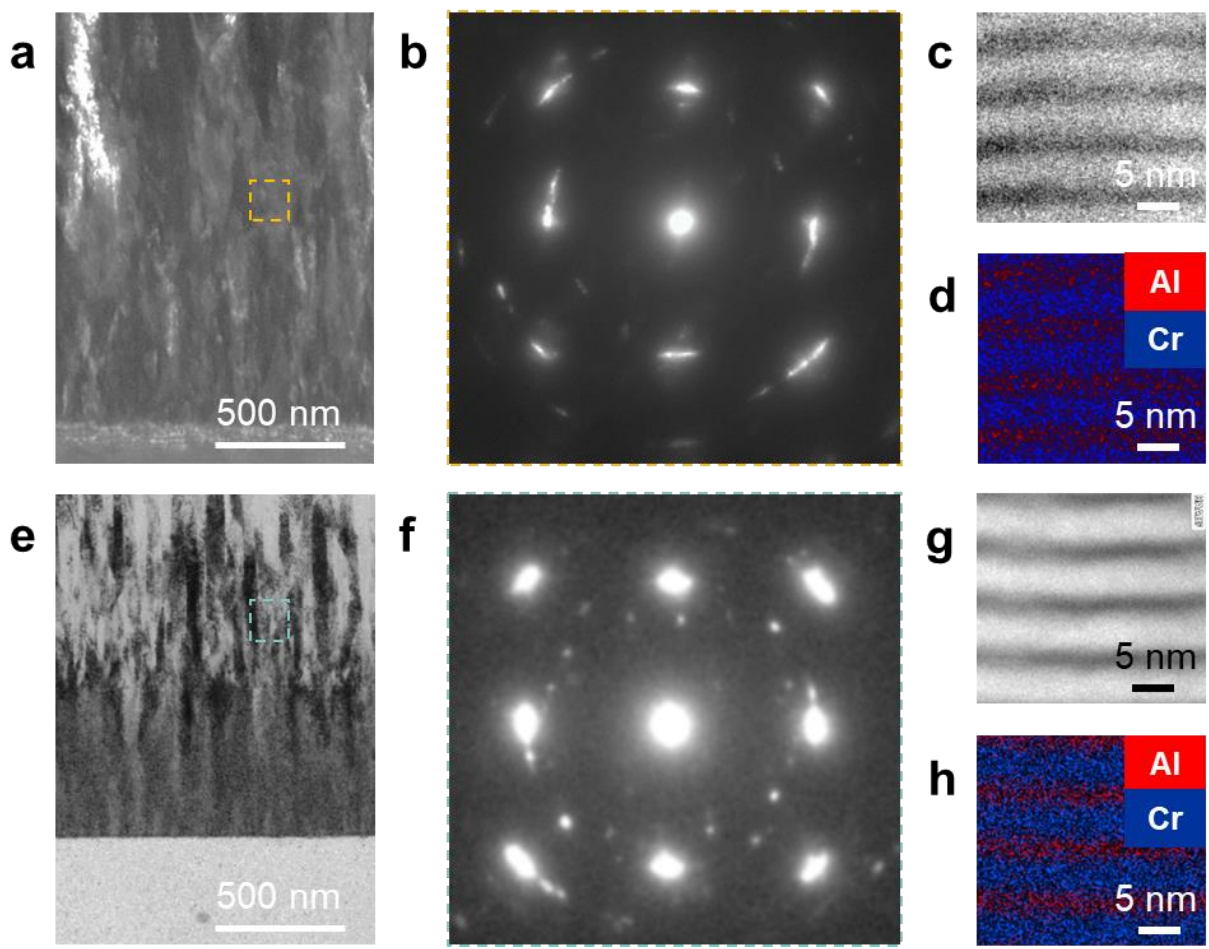

**Fig. S2** Microstructure and chemical analysis on the cg-coating (a–d) and cg/epi-coating (e–h). (a) and (e) TEM images of the cross-section of the coatings. (b) and (f) The selected area electron diffraction patterns from the columnar grain structures of each coating. (c) and (g) HAADF-STEM images of each coating for EDS mapping and corresponding maps are shown in (d) and (h), which display the multilayered structure, i.e., alternating CrN and AlN.

**Intergranular fracture surface of the cg and the cg/epi coating without FIB effect**

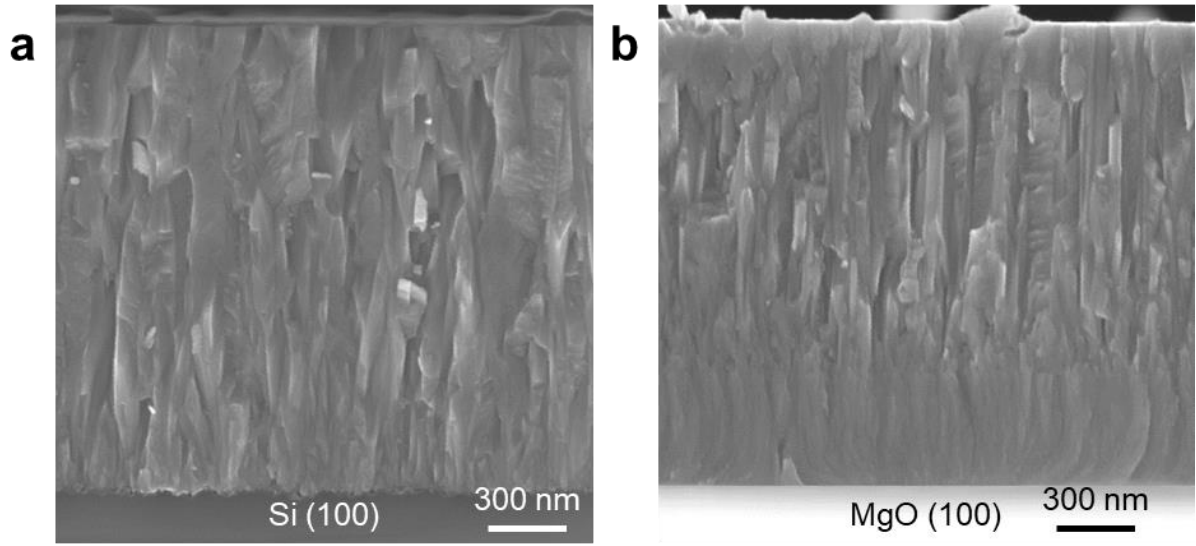

**Fig. S3** Fracture surfaces of (a) cg-coating and (b) the cg/epi-coating broken manually from the substrate side.